\documentclass[12pt]{article}
\usepackage[utf8]{inputenc}
\usepackage{amsmath, amssymb}
\usepackage{graphicx}
\usepackage{hyperref}
\usepackage[margin=1in]{geometry}
\usepackage{pgfplots}
\usepackage{pgfplotstable}
\usepackage{booktabs}
\usepackage{tikz}
\usepackage{array}
\pgfplotsset{compat=1.17}
\usepackage{subcaption}
\usepackage{caption}
\usepackage{authblk}
\usepackage{multicol}
\usepackage{enumitem}
\title{Optimal Classification of Three-Qubit Entanglement with Cascaded Support Vector Machine}

\author[]{Fatemeh Sadat Lajevardi}
\author[]{Azam Mani}
\author[]{Ali Fahim}
\affil[]{Department of Engineering Science, College of Engineering, University of Tehran, Iran}

\date{}

\begin{document}
\maketitle

\begin{abstract}
	We introduce a systematic framework for three-qubit entanglement classification using a cascaded architecture of Support Vector Machine (SVM) classifiers. Leveraging the well defined three-qubit structure with the four nested entanglement classes (S, B, W, and GHZ), we construct three distinct witness models ($\mathcal{M}_{B}$, $\mathcal{M}_{W}$, and $\mathcal{M}_{GHZ}$) that sequentially discriminate between these classes. The proposed Cascaded model achieves an overall classification accuracy of $95\%$ on a comprehensive dataset of mixed states. The framework's robustness and generalization capabilities are confirmed through rigorous testing against out-of-distribution (OOD) entangled states and various quantum noise channels, where the model maintains high performance. A key contribution of this research is an optimization protocol based on systematic feature importance analysis. This approach yields a tunable framework that significantly reduces the number of required features, while maintaining reliable model accuracy.
\end{abstract}

\section{Introduction}
\label{sec:introduction}

Entanglement stands as a cornerstone of quantum mechanics and a critical resource for emerging quantum technologies. While bipartite entanglement is well understood, the characterization of multipartite entanglement remains a formidable challenge due to the exponentially growing complexity of the state space \cite{gharibian2010strong}. Three-qubit systems represent the first and most fundamental step into this complexity, exhibiting a rich structure of entanglement classes that are inequivalent under local operations \cite{bennett1996concentrating, Wolfgang2000, Sabın2008}. The complete classification of these states is not merely of theoretical interest but a crucial prerequisite for harnessing them in quantum computation, communication, and metrology \cite{chitambar2019quantum, Borsten2009, Prakash2011}.	
	
A major challenge in this field is the experimental verification of entanglement. While analytical tools such as entanglement witnesses provide a rigorous framework for detection~\cite{PPT, Michal1996, Gühne2009}, they often face practical limitations, including non-optimality and susceptibility to misclassification, particularly for mixed states near class boundaries~\cite{roik2023physrev, Gurvits2003, Liu2022}. In recent years, machine learning has emerged as a powerful approach to address such complex classification problems in quantum physics. Techniques including deep neural networks~\cite{jaroSemiSupervised, julioDNN, naema2023} and Support Vector Machines (SVMs)~\cite{sanavio2023, greenwoodML} have revealed hidden patterns in the data, enabling state classification based on empirical learning rather than purely analytical methods.

However, many advanced methods—particularly deep learning approaches—tend to function as ``black boxes,'' achieving high accuracy at the expense of physical interpretability. This lack of interpretability presents a significant barrier to experimental realization, as a model that cannot be physically interpreted cannot easily be translated into a measurable witness, and one would need to perform full state tomography to obtain the model’s output. 

In this work, we bridge this gap by introducing a systematic framework for classifying three-qubit entanglement using Support Vector Machines (SVMs)~\cite{svm}. We exploit the intrinsic geometric correspondence between the SVM decision hyperplane and the structure of entanglement witnesses, to design a cascaded classification protocol. Our method consists of three distinct SVM-based models, which shows the nested convex structure of the three-qubit entanglement classes~\cite{acinClassification3q}. This cascaded design enables the progressive and unambiguous identification of a quantum state’s entanglement class.

The central contribution of this paper extends beyond high-accuracy classification. We introduce a model optimization protocol aimed at experimental feasibility, systematically reducing the number of required features (equivalently, required quantum measurements) from full state tomography (63 independent parameters) to a minimal, resource-efficient subset. This is achieved through a robust feature selection algorithm that quantifies the importance of each Pauli observable. We demonstrate that this method not only achieves state-of-the-art classification accuracy but also yields a tunable framework for constructing entanglement classification models that are both powerful and practically resource efficient. We have also validated our method through extensive numerical simulations, including performance tests under noisy conditions and against out-of-distribution entangled states, confirming both the robustness and generalization capability of our approach. It should also be noted that our proposed method also reports better classification accuracy than several deep learning-based approaches \cite{sanavio2023, jaroSemiSupervised, julioDNN, naema2023}, and beyond that, it successfully considers the entire space of three-qubit mixed states, unlike other works which also employ an SVM-based model, but are restricted to classification of specific states \cite{greenwoodML, Mahdian2025}.

%It should also be noted that our proposed method also reports better classification accuracy than several deep learning-based approaches. For instance, the study \cite{jaroSemiSupervised} which employs convolutional neural networks, provides a lower average accuracy of $96\%$ for its best results in detecting entangled states with positive partial transpose \textcolor{red}{is it for two qubit systems?}, compared to the $99.9\%$ accuracy of the best results in our work. Similarly, the study in \cite{julioDNN} has achieved the accuracy of $90\%$  despite utilizing a multi-layer perceptron network, and the work \cite{naema2023} with the same method reaches only $80\%$ for the accuracy in classifying three-qubit states. Furthermore, our method demonstrates superior performance compared to \cite{sanavio2023}, which has reported $95\%$ accuracy on mixed states using an SVM model. Beyond these, our approach successfully generalizes to the entire space of three-qubit mixed states, unlike \cite{greenwoodML} which also employs an SVM-based model, but is restricted to Werner-like GHZ and W states.

This paper is organized as follows. We begin by reviewing the theoretical foundations of three-qubit entanglement classes in Section~\ref{sec:characterizing}. We then formulate the entanglement classification problem and introduce our witness-based approach in Section~\ref{sec:ent_classification}, followed by a description of the dataset generation procedure in Section~\ref{sec:datasetGeneration}. The development and evaluation of the classifiers are presented in Section~\ref{sec:models}, while Section~\ref{sec:optimalWitness} introduces the model optimization framework. We conclude in Section~\ref{sec:discussion} with a comparative analysis and discussion of the results.

\section{3-qubit Entanglement Classes}
\label{sec:characterizing}

\begin{figure}[h]
  \centering
  \includegraphics[scale=0.3]{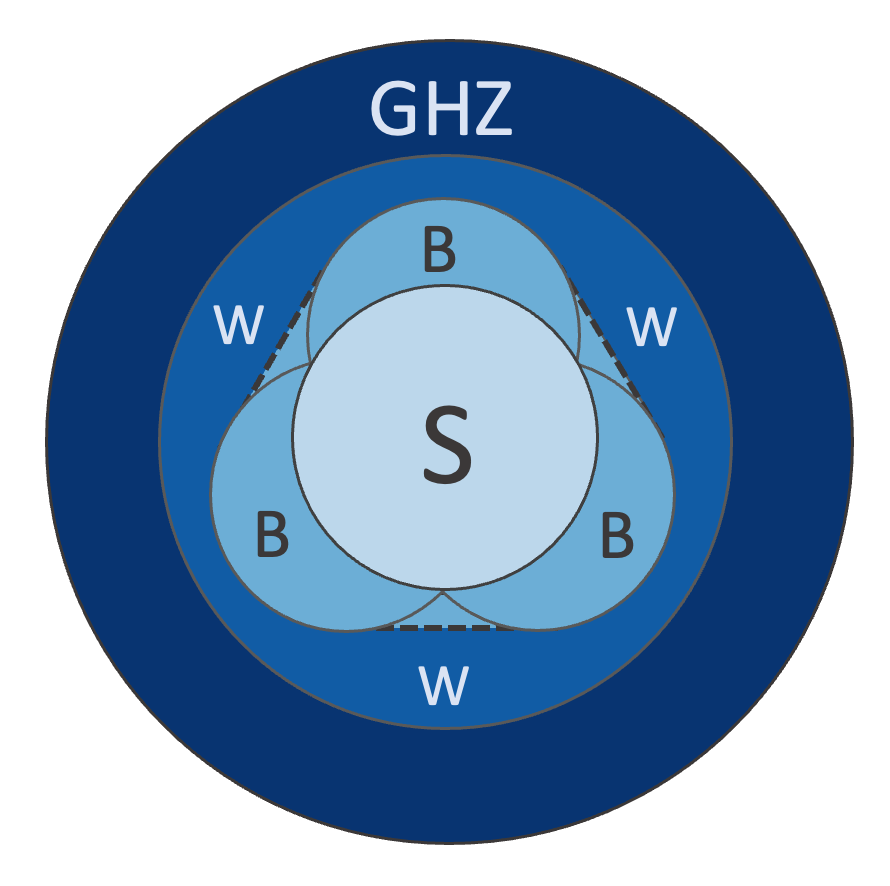}
  \caption{The schematic structure of three-qubit entanglement classes. $S$, $B$, $W$ and $GHZ$ refer to fully separable, biseparable, $W$ and $GHZ$ classes respectively. The four sets form a hierarchy structure $S \subseteq B \subseteq W \subseteq GHZ$.}
  \label{fig:structure}
\end{figure}

In \cite{acinClassification3q}, Acín et al. employed the Schmidt decomposition for three-party quantum systems to classify the entire space of mixed three-qubit states. A key feature of this classification is the compact and convex structure of the state space of each class, where these subspaces are hierarchically embedded within one another. Based on this theoretical groundwork, the space of three-qubit states can be divided to four locally-inequivalent classes; separable, biseparable, $W$-type and $GHZ$-type classes. While the $W$ and $GHZ$-type states are both entangled, they cannot be transformed into one another through local operations and classical communication. More precisely, any pure state admits a representation of the form of a $GHZ$-vector as follow:
\begin{equation}
|\psi_{GHZ}\rangle = \lambda_{0}|000\rangle + \lambda_{1} e^{i \mathit{\theta}}|100\rangle +\lambda_{2}|101\rangle + \lambda_{3}|110\rangle + \lambda_{4}|111\rangle.
\label{vectorGHZ_type}
\end{equation}
Where $\lambda_i \geq 0$, $\sum_i \lambda_i^2 = 1$ and $\theta \in [0, \pi]$. The notation $\{|0\rangle, |1\rangle\}$ represents an orthonormal basis of the Hilbert spaces of Alice, Bob or Charlie, whom we will henceforth refer to as A, B, and C for brevity.  
On the other hand, a $W$-type state can be written as
\begin{equation}
|\psi_{W}\rangle = \lambda_{0}|000\rangle + \lambda_{1}|100\rangle +\lambda_{2}|101\rangle + \lambda_{3}|110\rangle.
\label{vectorW_type}
\end{equation}
It is important to note that the standard GHZ and W states, characterized by the forms given in the following equations respectively, represent individual members of the $GHZ$ and $W$-type classes.

\begin{equation}
|GHZ\rangle = \frac{1}{\sqrt{2}} (|000\rangle + |111\rangle)
\label{vectorGHZ}
\end{equation}

\begin{equation}
|W\rangle = \frac{1}{\sqrt{3}} (|001\rangle + |010\rangle + |100\rangle)
\label{vectorW}
\end{equation}
%This characterization of three-qubit systems based on the Schmidt decomposition, serves as the foundation for the current work. 
According to this framework, mixed three-qubit systems can be partitioned into four convex and compact sets, as follow and detailed in \cite{brubCharEnt}. Figure \ref{fig:structure} illustrates this classification.

\begin{itemize}

\item \textbf{Class S}: Fully separable states are defined as convex combinations of pure product states $\lvert \psi_i\rangle$, i. e.
\begin{equation}
\rho_s = \sum_{i=1}^{m} \alpha_i  \ \lvert \psi_i \rangle \langle \psi_i \rvert 
\label{mixSep}
\end{equation}
where $m$ is an arbitrary positive integer and $\alpha_i$ are real, non-negative coefficients satisfying $\sum_{i=1}^{m} \alpha_i = 1$.

\item \textbf{Class B}: This class is constructed by mixing the fully separable states with at least one three-qubit state with bipartite entanglement, that is the  entangled states of the form $\rho_{AB} \otimes \rho_C$, $\rho_A \otimes \rho_{BC}$, or $\rho_B \otimes \rho_{AC}$. This explicitly excludes tripartite entanglement.

\item \textbf{Class W}: This  is formed by introducing at least one genuine three-qubit $W$-type state into the convex hull of class $B$.

\item \textbf{Class GHZ}: Similarly, this  class arises when at least one genuine $GHZ$-type state is included in the mixture of $W$ class states.

\end{itemize}

Under this classification scheme, the $GHZ$-type pure states reside exclusively in the set $GHZ \setminus W$, while the $W$-type pure states belong to $W \setminus B$. These four sets form a nested hierarchy expressed as $S \subseteq B \subseteq W \subseteq GHZ$, where each set is by construction convex. Given these structural properties, the classification of states into these sets becomes tractable.

\section{Entanglement Classification}
\label{sec:ent_classification}

In three-qubit systems, each quantum state belongs to one of four distinct entanglement classes, rendering their analysis a classification problem. 
The principal analytical tools which provide sufficient conditions for class memberships, are known to be entanglement witnesses. A quantum witness, formally represented as a Hermitian operator $\mathcal{W}$, serves to discriminate between the states of two sets $\mathcal{S}_1$ and $\mathcal{S}_2 \setminus \mathcal{S}_1$, respectively without and with a specific property, through the satisfaction of the inequalities
\begin{equation}
	\left\{
	\begin{aligned}
		&\mathrm{Tr}(\mathcal{W} \rho) \geq 0 \quad && \text{for all states without the desired property, i. e. }\forall \rho \in \mathcal{S}_1, \\
		&\mathrm{Tr}(\mathcal{W} \rho) < 0 \quad && \text{for at least one }\rho \in \mathcal{S}_2 \setminus \mathcal{S}_1.
	\end{aligned}
	\right.
	\label{eq:ent_witness}
\end{equation}
A representative example is the operator
\begin{equation}
	\mathcal{W}_{GHZ} = \tfrac{3}{4} I - P_{GHZ}, \quad
	P_{GHZ} = |GHZ\rangle \langle GHZ|,
	\label{eq:ghz_witness}
\end{equation}
introduced in \cite{acinClassification3q}, for discriminating between the sets $W$ and $GHZ \setminus W$, where $|GHZ\rangle$ denotes the standard GHZ state defined in Eq.~(\ref{vectorGHZ}). 
A negative expectation value $\mathrm{Tr}(\mathcal{W}_{GHZ}\,\rho)<0$ certifies that the state $\rho$ belongs to the class $GHZ \setminus W$. 
Conversely, a non-negative value $\mathrm{Tr}(\mathcal{W}_{GHZ}\,\rho) \ge 0$ cannot conclusively establish membership in $W$, since some states in $GHZ \setminus W$ may also yield positive expectation values. 
Similarly, the witness
\begin{equation}
	\mathcal{W}_{W} = \tfrac{2}{3} I - P_{W}, \quad
	P_{W} = |W\rangle \langle W|,
	\label{eq:w_witness}
\end{equation}
with $|W\rangle$ being the standard $W$ state defined in Eq.~(\ref{vectorW}), is also introduced in \cite{acinClassification3q} to identify membership in $W \setminus B$. 
As before, $\mathrm{Tr}(\mathcal{W}_{W}\,\rho) < 0$ indicates that $\rho$ belongs to $W \setminus B$, while $\mathrm{Tr}(\mathcal{W}_{W}\,\rho) \ge 0$ remains inconclusive. 
To reduce such inconclusive regions, recent studies have focused on optimizing witness operators so that their decision boundaries align more closely with the true class frontier.\\

\begin{figure}[h]
	\centering
	\includegraphics[scale=0.25]{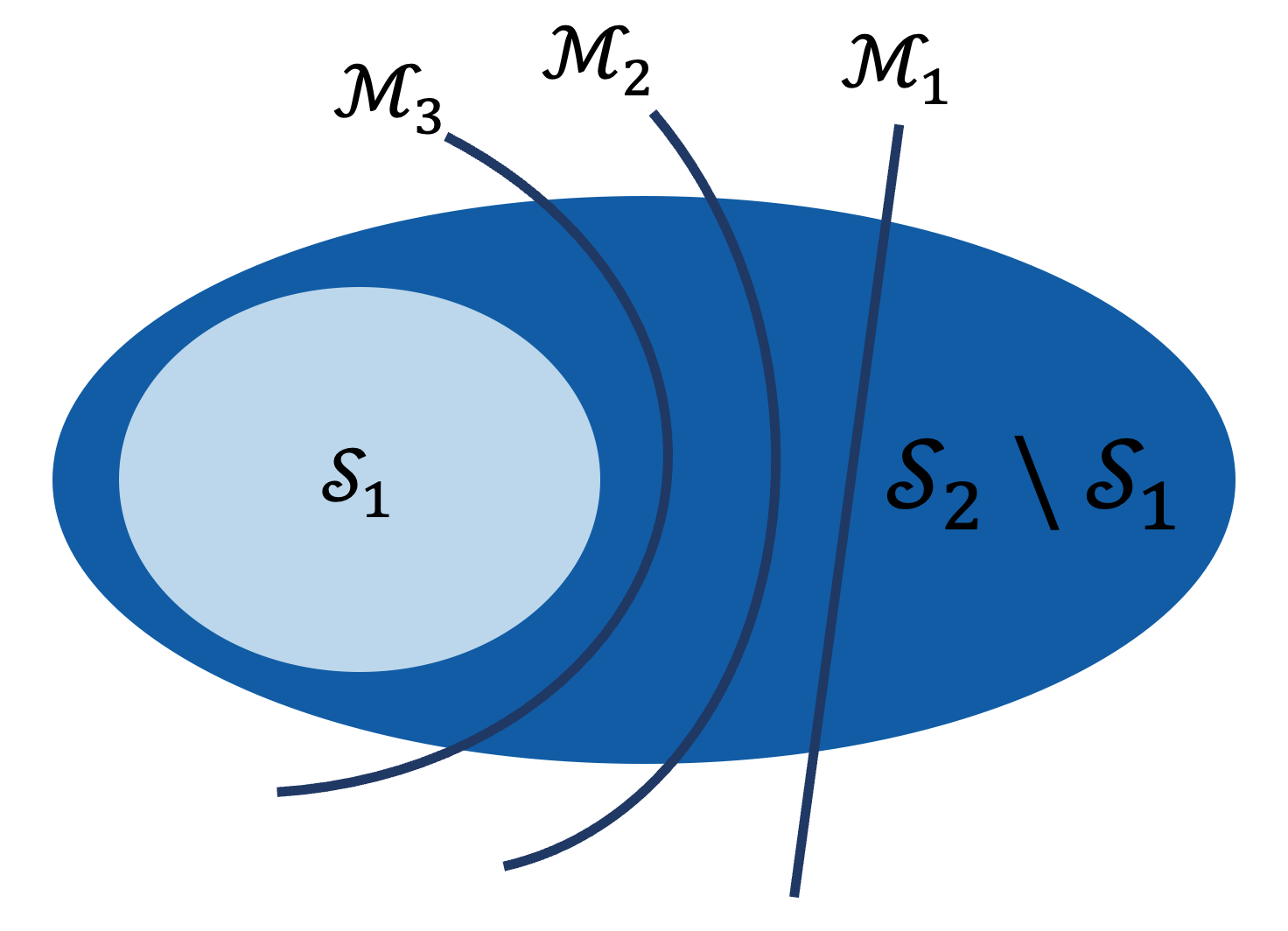}
	\caption{The schematics of three distinct witness models, 
		$\mathcal{M}_{1}$, $\mathcal{M}_{2}$, and $\mathcal{M}_{3}$, 
		all constructed to discriminate between the sets $\mathcal{S}_1$ and $\mathcal{S}_2 \setminus \mathcal{S}_1$. Among these, $\mathcal{M}_{3}$ demonstrates a higher degree of optimality, attaining improved accuracy by positioning its decision plane in closer proximity to the boundary of $\mathcal{S}_1$.}
	\label{fig:witness}
\end{figure}

Recent advances include machine learning-based classifiers, which have demonstrated remarkable efficiency in discriminating three-qubit entanglement classes \cite{chenUnsupervise, greenwoodML, jaroSemiSupervised, julioDNN}. The Support Vector Machine (SVM) stands out as one of the most effective machine learning techniques for the construction of entanglement witness models. Figure \ref{fig:witness} illustrates the structure of a witness model constructed using this method, highlighting its capacity to separate two classes and optimize the overall performance of the model.

Support Vector Machines (SVMs) demonstrate particular promise for entanglement detection due to their geometric properties. This suitability stems from the fundamental similarity between the mathematical structure of quantum witnesses (Eq. (\ref{eq:ent_witness})) and SVM decision boundaries. The core objective of SVM is to find a hyperplane $M^T \mathbf{x} + b = 0$, where $\mathbf{x} \in R^n$ represents the $n$-dimensional input feature vector, $M$ denotes the normal vector to the hyperplane, and  $b$ is the bias term. The SVM optimization aims to find the maximal margin hyperplane that separates two distinct classes (denoted here as class 1 and class 2):

\begin{equation}
	\left\{
	\begin{aligned}
		& M^T \mathbf{x} + b \geq +1 \quad && \text{for class 1,} \\
		& M^T \mathbf{x} + b \leq -1 \quad && \text{for class 2.}
	\end{aligned}
	\right.
	\label{eq:svm_constraints}
\end{equation}

Consequently, SVMs provide a rigorous machine-learning framework for the systematic construction of entanglement witnesses. In this setting, the optimized separating hyperplane obtained from the SVM can be directly mapped to a corresponding witness operator \cite{Havlicek2019}. The key distinction between machine-learning-based witnesses and analytically constructed ones lies in the superior optimality achieved by the former. In particular, machine-learning-derived witnesses significantly reduce the number of inconclusive cases and, within a specified precision, enable the unambiguous assignment of each quantum state to a definite entanglement class.\\

In this study, however, we do not rely on this explicit construction and instead emphasize the underlying conceptual framework of SVMs. When considered in their entirety, any SVM model formulated for entanglement classification may be interpreted as an entanglement-witness model (rather than a physical witness operator). With this perspective, we adopt an SVM-based framework to construct independent witness models capable of distinguishing among the four entanglement classes in three-qubit systems. The compactness and convexity of the quantum state space provide a natural and robust foundation for implementing such constructions. A complete characterization of the system therefore requires three independent witness models:

\begin{itemize}
	
	\item $\mathcal{M}_{GHZ}$: Model to certify the membership in  $GHZ \setminus W$,
	
	\item $\mathcal{M}_{W}$: Model to certify the membership in $W \setminus B$,
	
	\item $\mathcal{M}_{B}$: Model to certify the membership in  $B \setminus S$.
	
\end{itemize} 

\begin{figure}[h]
    \centering
    \includegraphics[scale=0.4]{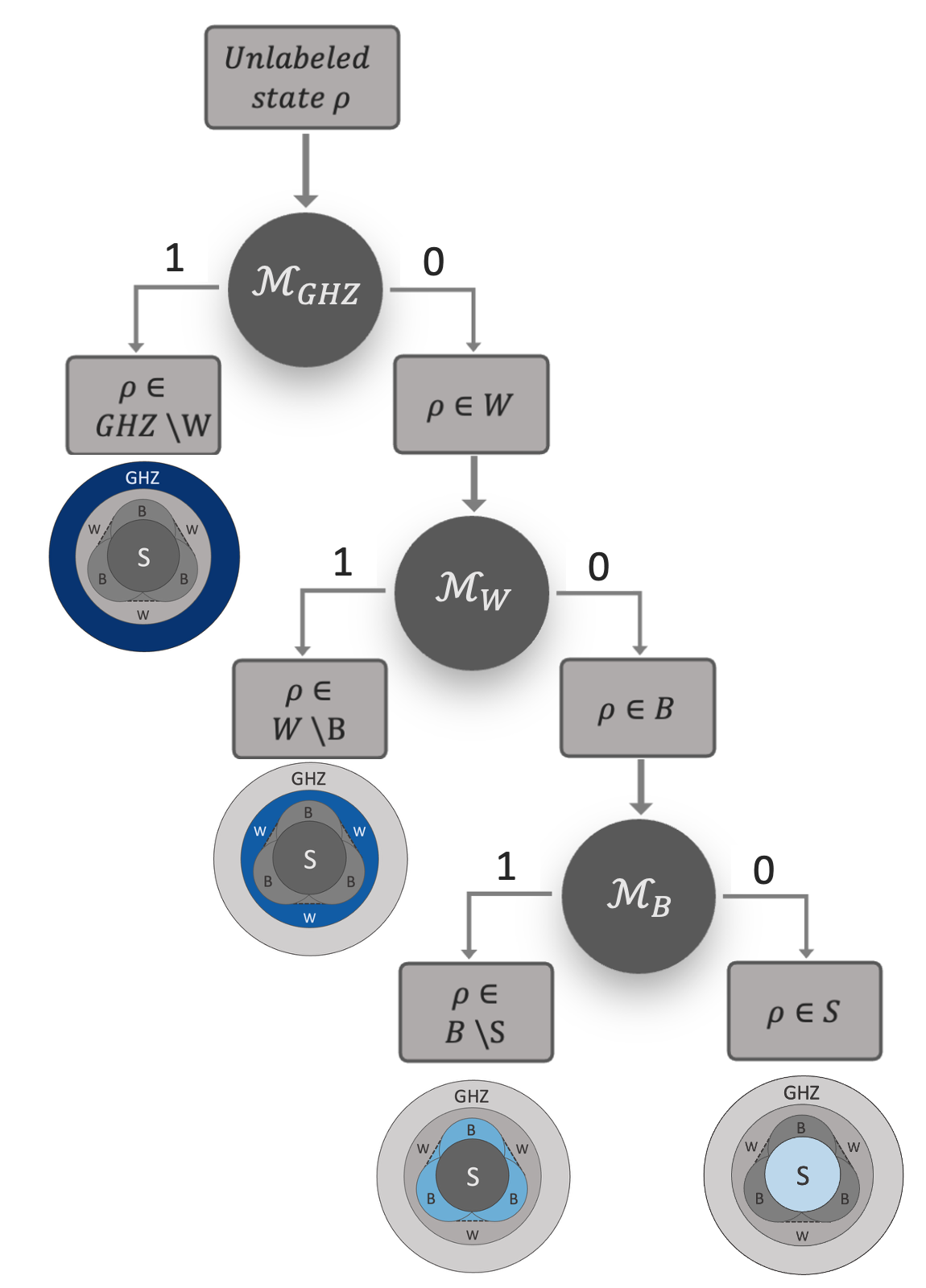}
    \caption{Cascaded classification of an unlabeled quantum state $\rho$ using a sequence of three models. The process begins with $\mathcal{M}_{GHZ}$: the state $\rho$ may be classified as belonging to $GHZ \setminus W$; otherwise, it is assigned to the $W$ category. In this latter case, the subsequent application of $\mathcal{M}_{W}$ distinguishes between $W \setminus B$ and $B$. Finally, for states identified within $B$, the model $\mathcal{M}_{B}$ determines whether $\rho \in B \setminus S$ or $\rho \in S$. This cascaded framework achieves precise and interpretable classification through sequential hyperplane-based decision boundaries.}
    \label{fig:modelStructure}
\end{figure}

The cascaded classification protocol employs a cascade of SVM models to systematically determine the membership of a quantum state within the entanglement hierarchy. The process begins with the model $\mathcal{M}_{GHZ}$, which distinguishes states belonging to the set $GHZ \setminus W$. States not identified within this category are subsequently analyzed by $\mathcal{M}_{W}$, which resolves the distinction between $W \setminus B$ and $B$. Finally, for states outside $W \setminus B$, the model $\mathcal{M}_{B}$ provides the final discrimination between $B$ and $S$, corresponding to the biseparable and fully separable classes, respectively. This tiered scheme incrementally eliminates classification ambiguities at each stage, yielding a complete and operationally efficient characterization of three-qubit entanglement. The full decision pathway is depicted in Figure~\ref{fig:modelStructure}.\\

Beyond the cascaded architecture itself, the key contribution of our framework lies in the optimized construction of these models, which enhances practical applicability by minimizing the number of required features while preserving discriminative power. As discussed in Section~\ref{sec:optimalWitness}, this optimization confers a significant advantage over existing approaches \cite{chenUnsupervise, julioDNN, jaroSemiSupervised, greenwoodML}.

\section{Dataset Generation}
\label{sec:datasetGeneration}

In this section, we describe the procedure for generating the datasets required to construct each model. The preparation of these datasets, together with the appropriate labeling, constitute necessary steps for developing machine-learning-based entanglement witnesses, which will be further discussed in Section~\ref{sec:optimalWitness}. Before presenting the details of dataset construction, we first clarify the notation and representation used for the feature vectors. Any three-qubit density matrix can be expanded in terms of the identity and Pauli operators as
\begin{equation}
	\rho = \frac{1}{8}
	\sum_{i,j,k=0}^{3} t_{ijk} \ \sigma_i \otimes \sigma_j \otimes \sigma_k \ ,
	\label{eq:pauli}
\end{equation}
where $\otimes$ denotes the tensor product, $\sigma_0 = I$ is the identity operator, $\sigma_i$ $(i \in \{1,2,3\})$ are the Pauli matrices, and $t_{ijk}$ are real parameters. This representation provides a more physically interpretable framework than the purely algebraic description in terms of density matrix elements. \\

Using this expansion, any arbitrary three-qubit state can be represented by a  real $63$-dimensional feature vector $\mathbf{x}$ composed of the coefficients $t_{ijk} = \mathrm{Tr}(\rho\, \sigma_i \otimes \sigma_j \otimes \sigma_k)$. Each coefficient $t_{ijk}$ corresponds to the expectation value of the Hermitian operator $\sigma_i \otimes \sigma_j \otimes \sigma_k$, and can therefore be experimentally obtained through local measurements of the associated operators.
Although the identity and Pauli matrices are conventionally denoted as $\sigma_0$, $\sigma_1$, $\sigma_2$, and $\sigma_3$, it is sometimes convenient to use their equivalent notation $I$, $X$, $Y$, and $Z$, respectively. Accordingly, compact forms such as $XIX$ and $XYZ$ are used to represent $\sigma_1 \otimes \sigma_0 \otimes \sigma_1$ and $\sigma_1 \otimes \sigma_2 \otimes \sigma_3$, respectively, with analogous notation applied to other combinations.

\subsection{Dataset for $\mathcal{M}_{B}$ }

To construct the model separating class $S$ from $B \setminus S$, we generate datasets for both categories, assigning label $0$ to states in $S$ and label $1$ to states in $B \setminus S$. Class $S$ comprises convex combinations of fully separable states, which are naturally represented as tensor products of three single-qubit density matrices:
\begin{equation}
	\rho_{\text{sep}} = \rho_1 \otimes \rho_2 \otimes \rho_3,
\end{equation}
where each $\rho_i$ is sampled independently from a suitable distribution.\\

Pure single-qubit states are drawn from the Haar measure, providing a uniform distribution over the Bloch sphere. In practice, Haar-random unitaries $U \in U(2)$ are generated using QuTiP’s \texttt{rand\_unitary\_haar} function, yielding states
\begin{equation}
	\rho_{\psi} = |\psi\rangle \langle \psi|, \qquad |\psi\rangle = U |0\rangle,
\end{equation}
where $U \sim \text{Haar}(U(2))$, and $|0\rangle$ is the conventional vector of the computational basis.\\

Mixed states are generated following the Mai--Alquier (MA) distribution~\cite{MADistribution}, which constructs density matrices as convex combinations of Haar-distributed pure states,
\begin{equation}
\rho_{\text{MA}} = \sum_{i=1}^N \alpha_i \, |\psi_i\rangle \langle \psi_i|,
\end{equation}
where $|\psi_i\rangle$ are independent Haar-random pure states on the underlying Hilbert space, $N \in \{1,\dots,50\}$ is the number of components, and the weights $\boldsymbol{\alpha} = (\alpha_1,\ldots,\alpha_N)$ form a random probability vector. The weights are drawn from a Dirichlet distribution,
\begin{equation}
(\alpha_1, \ldots, \alpha_N) \sim \operatorname{Dir}(\beta_1, \ldots, \beta_N),
\qquad
\sum_{i=1}^{N} \alpha_i = 1,\;\; \alpha_i \ge 0,
\end{equation}
where the Dirichlet parameters $\boldsymbol{\beta} = (\beta_1,\ldots,\beta_N)$ are positive real numbers that control the concentration of the distribution over the simplex, with larger $\beta_i$ yielding more balanced mixtures and smaller $\beta_i$ producing sparser mixtures. This methodology ensures that all generated states are valid density matrices—positive semidefinite with unit trace—while providing broad coverage of separable and mixed states. Although tensor products of Haar-random single-qubit states do not produce a uniform distribution across the full multipartite Hilbert space, this structured sampling preserves reproducibility and yields datasets suitable for robust machine-learning training.\\

To generate the set $B \setminus S$ (label $1$), corresponding to biseparable states, we extend the previous construction by combining single-qubit states with two-qubit entangled states. Specifically, a biseparable state is formed as the tensor product of a single-qubit density matrix, prepared as described above, and a two-qubit entangled state. To obtain a representative and diverse ensemble of two-qubit entangled states, Monte Carlo sampling is employed: random two-qubit states are generated and subsequently tested for entanglement using the Peres–Horodecki Positive Partial Transpose (PPT) criterion~\cite{PPT}. States that pass this test are then paired with single-qubit states via the tensor product to form biseparable states, ensuring coverage of all bipartitions. Formally, the resulting biseparable states are constructed as
\begin{equation}
	\rho_{B \setminus S}
	=
	\sum_{i=1}^{N} \gamma_i\,
	\big( \rho^{(i)}_{L_i} \otimes \rho^{(i)}_{M_i N_i} \big),
	\qquad
	\gamma_i \ge 0,\quad \sum_{i=1}^{N}\gamma_i = 1.
	\label{eq:biSep}
\end{equation}
Where, for each index $i$, the ordered pair $(L_i, M_i N_i)$ corresponds to one of the three possible bipartitions $\{(A,BC), (B,AC), (C,AB)\}$. Here, $\rho^{(i)}_{L_i}$ denotes a  single-qubit density matrix and $\rho^{(i)}_{M_i N_i}$ represents a two-qubit entangled state.
This construction ensures that the generated biseparable states respect the proper bipartition structure while preserving the statistical diversity inherited from both single- and two-qubit components, thereby providing a consistent and reproducible dataset for training machine-learning-based entanglement witnesses.

\subsection{Dataset for $\mathcal{M}_{W}$  }
The construction of this model relies on two datasets; one corresponding to class $B$, here labeled as $0$, and the other to the complementary set $W \setminus B$, labeled as $1$.
The dataset associated with class $B$ is generated by combining and mixing the two datasets obtained in the previous step, namely $B \setminus S$ and $S$. Irrespective of their original labels, all states within this union are here relabeled as $0$, since their entirety lies within the region $B$, which itself is interior to $W \setminus B$. To ensure comprehensive coverage of class $B$, the dataset is supplemented with the convex combinations of individual samples drawn from both $B \setminus S$ and $S$. Formally, this construction can be expressed as
\begin{equation}
	B = \mathrm{convex\ hull}\{ (B \setminus S) \cup S \}.
	\label{eq:classB}
\end{equation}

To generate samples for the class $W \setminus B$, labeled as $1$, rotated versions of the standard $W$ state~\eqref{vectorW} are employed. Random local unitary operators are applied to the standard $W$ state, thereby producing a diverse ensemble of states that remain within the same entanglement class, as local unitaries preserve the entanglement type.
Moreover, to achieve better coverage of the data, additional samples between $B$ and $W \setminus B$ are incorporated~\cite{polson2011data}. These states include locally rotated versions of the Werner–$W$ state, which is a convex mixture of the standard $W$ state and the maximally mixed state,
\begin{equation}
	\rho_{\mathrm{WW}} = \alpha \, |W\rangle\langle W| + (1-\alpha)\, 
	\frac{I^{\otimes 3}}{8},
	\quad \alpha \in [0,1],
	\label{eq:mixWandB}
\end{equation}
where $\alpha$ denotes the mixing parameter and $I^{\otimes 3}$ is the three-qubit identity operator. 
For the mixing parameter $\alpha$, there exists a critical threshold $\alpha_c =\frac{3}{7}$ such that \cite{siewert2011}:

\begin{itemize}
\item For $\alpha > \alpha_c$, the state belongs to the class $W \setminus B$,
\item For $\alpha \leq \alpha_c$, the state belongs to $B$.
\end{itemize}

By generating states for both classes based on this criterion and introducing locally rotated versions of~\eqref{eq:mixWandB} into dataset, we enhance that's diversity and training value.

\subsection{Dataset for $\mathcal{M}_{GHZ}$}

To derive a machine-learning-based witness model capable of distinguishing between the classes $\mathrm{GHZ} \setminus W$ and $W$ ($\mathcal{M}_{GHZ}$), we follow the same methodology as in the previous construction. The $W$ class is formed by combining and intermixing two subsets of the dataset associated with the model $\mathcal{M}_{W}$—specifically, the sets $W \setminus B$ and $B$—both of which collectively constitute label~0 in the new dataset. In contrast, the $\mathrm{GHZ} \setminus W$ class is represented by standard GHZ state (\ref{vectorGHZ}) along with its randomly locally rotated counterparts, all assigned to label~1.\\

To enhance the dataset diversity, we also use the states of a mixture of the two classes as 

\begin{equation}
\rho_1 = (1 - \varepsilon) |\mathrm{GHZ}\rangle \langle \mathrm{GHZ}| + \varepsilon |\mathrm{W}\rangle \langle \mathrm{W}|,
\label{eq:mixWandGHZ}
\end{equation}

where $\varepsilon \in [0,1]$ is a real-valued mixing parameter. The state $\rho_1$ exhibits a critical threshold $\varepsilon_c$ that determines its entanglement class:

\begin{itemize}
\item For $\varepsilon \geq \varepsilon_c$, $\rho_1$ belongs to the $W$ class.
\item For $\varepsilon < \varepsilon_c$, $\rho_1$ falls into the $\mathrm{GHZ}\setminus W$ category.
\end{itemize}

The threshold is derived analytically by examining the entanglement structure of the system via partial trace operation. For genuinely tripartite GHZ entanglement, tracing out any one subsystem yields a separable bipartite state. By contrast, if the reduced state is found to be entangled—according to the positive partial transpose (PPT) criterion—the original entanglement must have been pairwise rather than genuinely GHZ-type. This distinction leads to a critical threshold of $\varepsilon_c =0.708$ \cite{Eltschka2008}, which provides a sharp decision boundary for generating dataset of both labels.

\section{Classifiers Models}
\label{sec:models}

\begin{figure}[h!]
\centering
\begin{subfigure}[b]{0.3\textwidth}
    \centering
    \includegraphics[width=\textwidth]{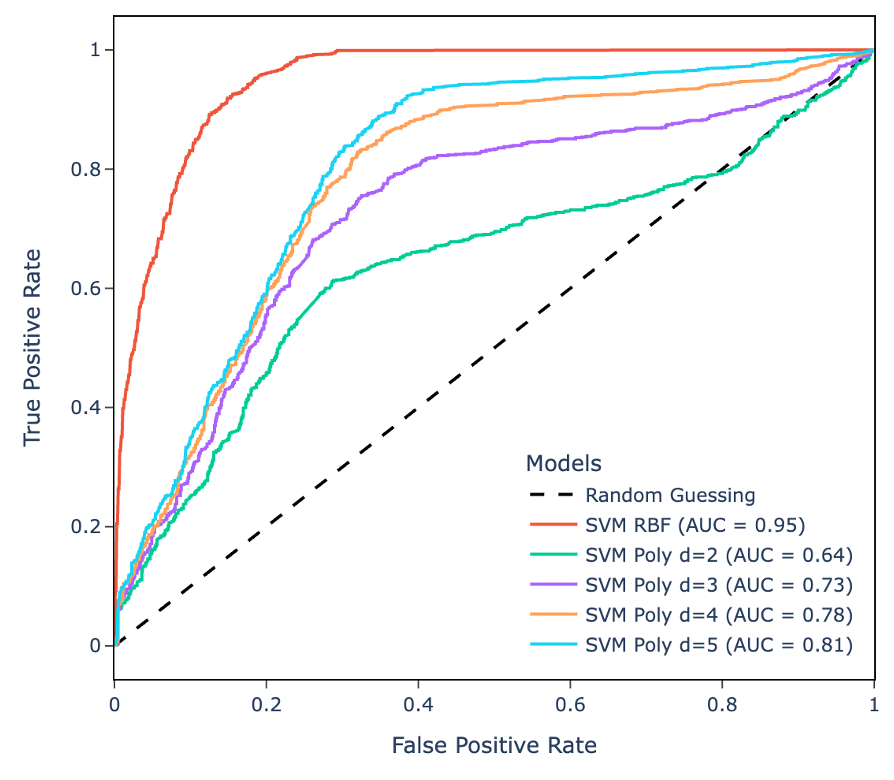}
    \caption{$\mathcal{M}_{GHZ}$}
\end{subfigure}
\hfill
\begin{subfigure}[b]{0.3\textwidth}
    \centering
    \includegraphics[width=\textwidth]{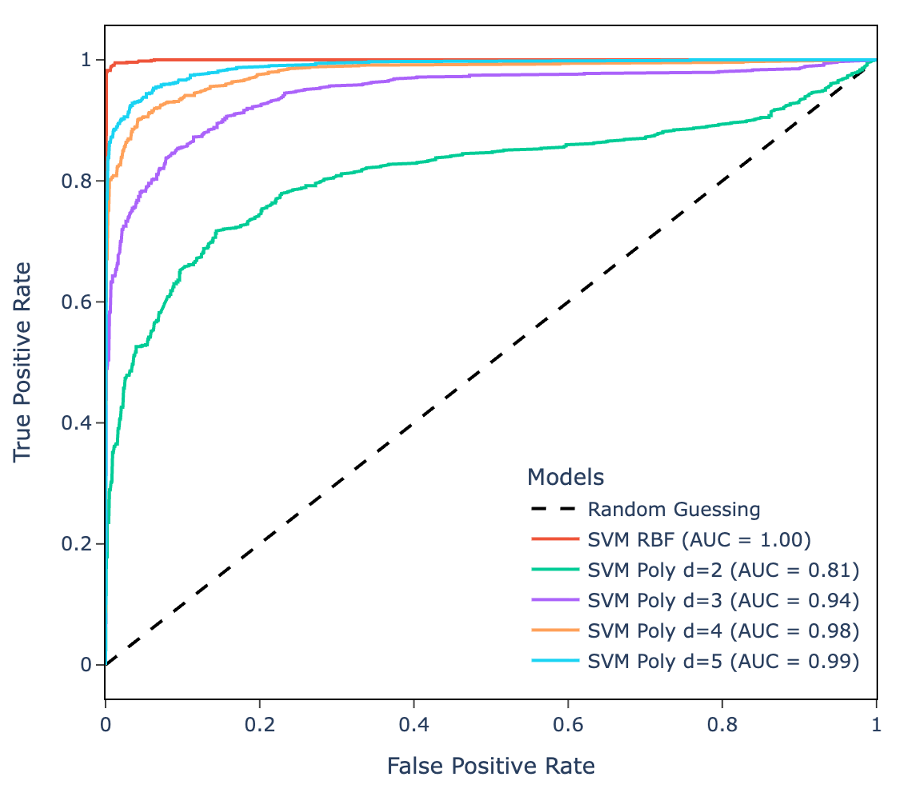}
    \caption{$\mathcal{M}_{W}$}
\end{subfigure}
\hfill
\begin{subfigure}[b]{0.3\textwidth}
    \centering
    \includegraphics[width=\textwidth]{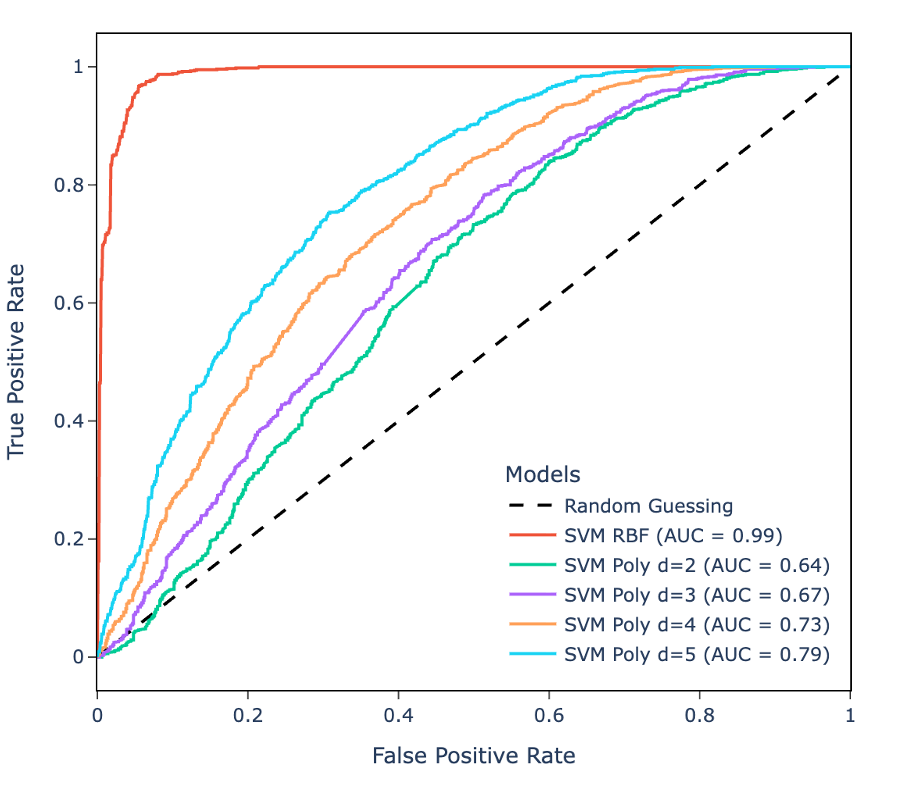}
    \caption{$\mathcal{M}_{B}$}
\end{subfigure}

\vspace{1em}
\caption{ROC curves for the three witness models. Each panel displays the performance of soft-margin SVM classifiers employing either a RBF kernel or polynomial kernels of degrees 2–9. Across all models, the RBF kernel (red curves) consistently achieves superior discrimination, exhibiting higher true positive rates over the entire range of false positive rates when compared with the best-performing polynomial kernels.}
\label{fig:poly_degree_tuning}
\end{figure}

\begin{figure}[h!]
\centering
\begin{subfigure}[b]{0.3\textwidth}
    \centering
    \includegraphics[width=\textwidth]{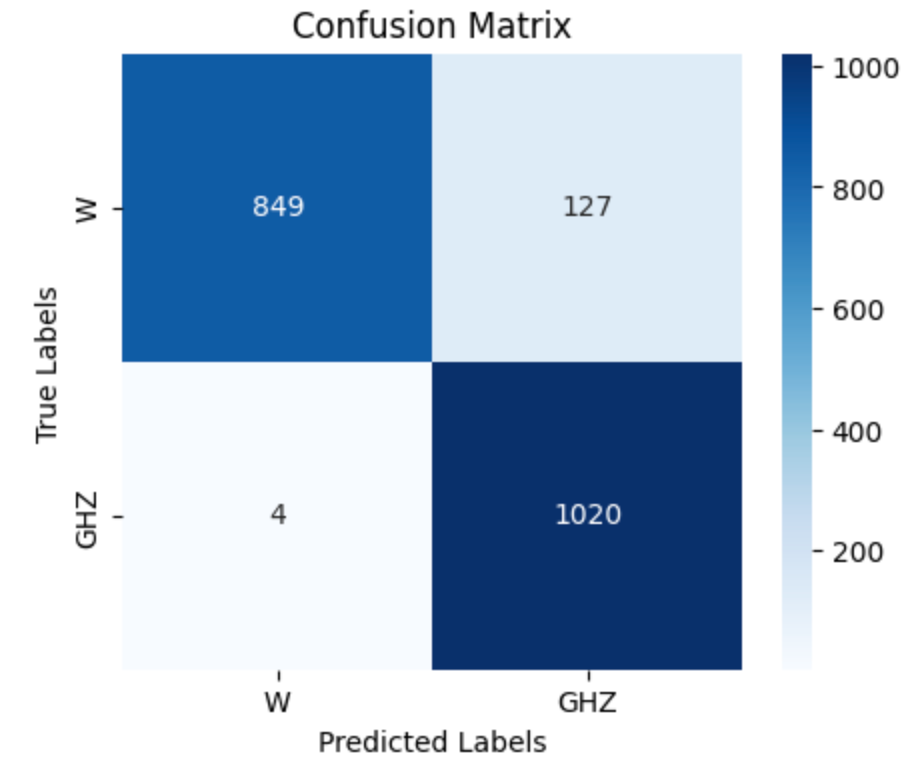}
    \caption{$\mathcal{M}_{GHZ}$}
\end{subfigure}
\hfill
\begin{subfigure}[b]{0.3\textwidth}
    \centering
    \includegraphics[width=\textwidth]{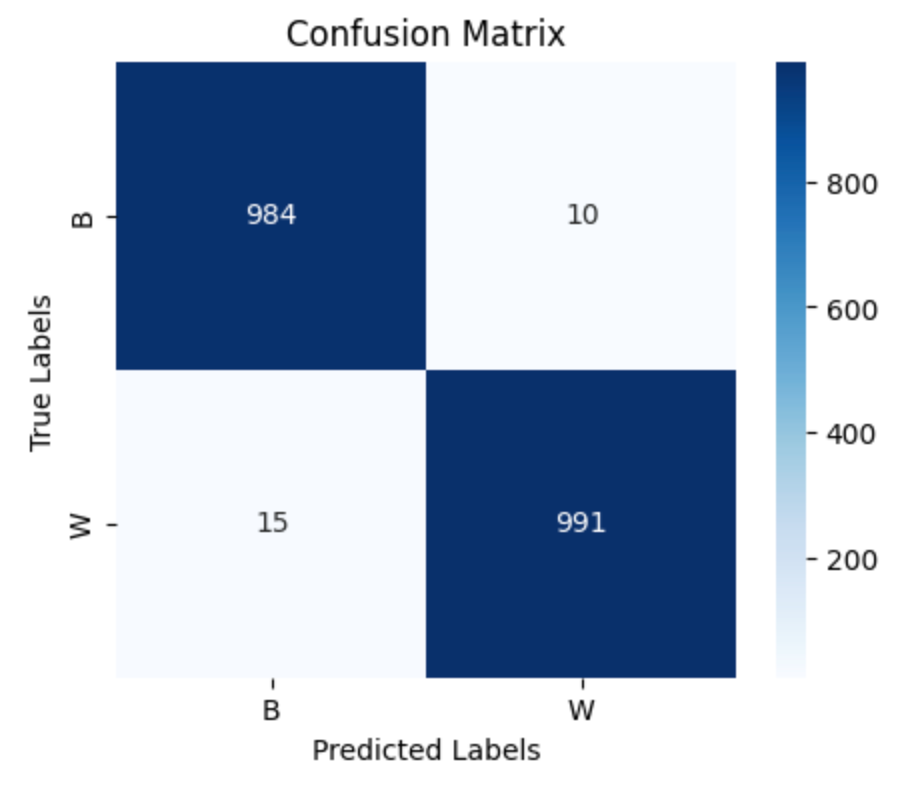}
    \caption{$\mathcal{M}_{W}$}
\end{subfigure}
\hfill
\begin{subfigure}[b]{0.3\textwidth}
    \centering
    \includegraphics[width=\textwidth]{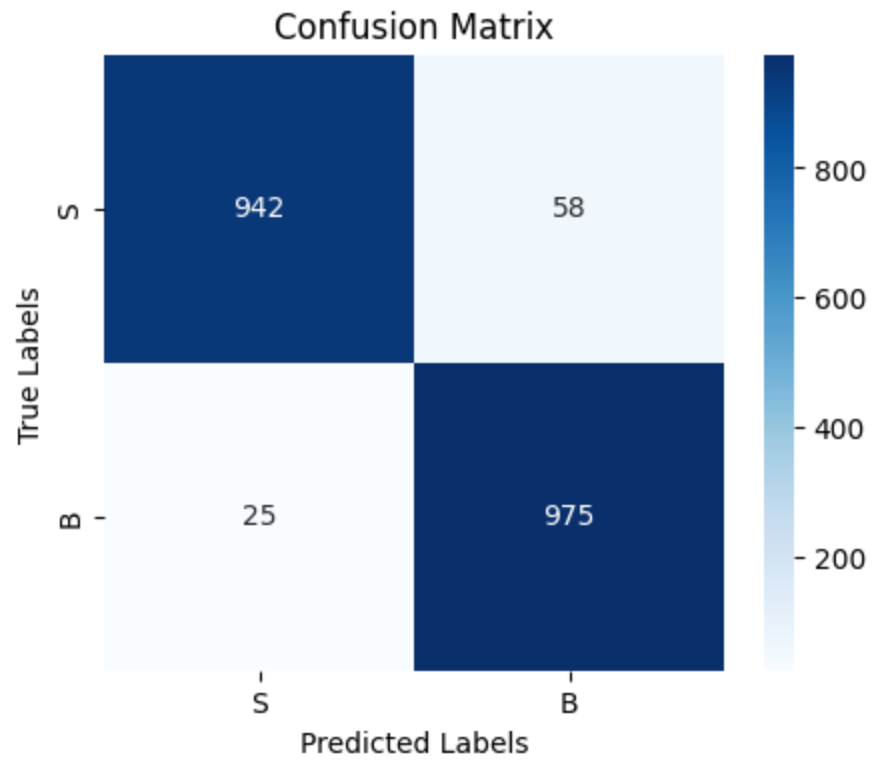}
    \caption{$\mathcal{M}_{B}$}
\end{subfigure}

\vspace{1em}
\caption{Confusion matrices for the three witness models. Each matrix compares predicted and true labels. Diagonal entries correspond to correct classifications, while off-diagonal entries indicate misclassifications between the entanglement classes.}
\label{fig:confiusionMatrix}
\end{figure}

\begin{figure}[h]
    \centering
    \includegraphics[scale=0.3]{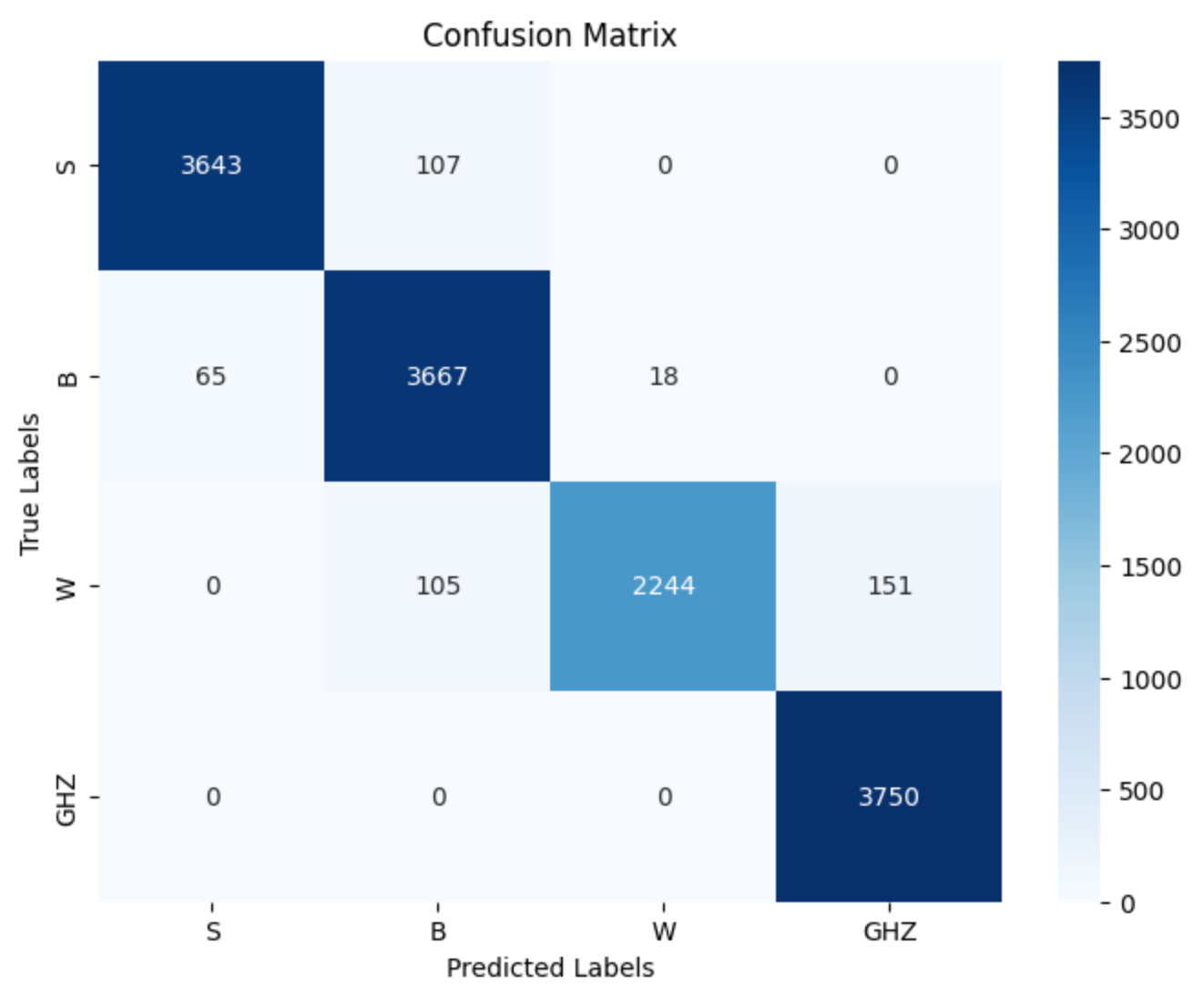}
    \caption{Confusion matrices associated with sequentially combination of all models, with each state classified into one of the sets: $S$, $B \setminus S$, $W \setminus B$, and $GHZ \setminus W$.}
    \label{fig:CMHoleModel}
\end{figure}

The initial step in developing the models discussed in the previous section involves the preparation of datasets. Three distinct datasets were generated, each consisting of 100,000 quantum states with binary classification labels (0 and 1) corresponding to each model. Each dataset was partitioned into three subsets for model development and evaluation: 70\% for training, 15\% for validation, and 15\% for testing. 
For all three models, we employed soft-margin Support Vector Machine (SVM) classifiers with kernel functions. Kernel selection was performed through a systematic analysis based on receiver operating characteristic (ROC) curves. Among the kernels considered, the radial basis function (RBF) kernel consistently outperformed polynomial kernels of degrees 2–9 (see Figure~\ref{fig:poly_degree_tuning}). This superiority is reflected in higher areas under the ROC curve (AUC) and more favorable ROC profiles across all kernels, with AUC values approaching unity.  Based on this robust performance, the RBF kernel was selected for all models, with its hyperparameters carefully tuned individually for each model by using the validation subset of the data.
The mathematical form of the RBF kernel is given by:

\begin{equation}
K(\mathbf{x},\mathbf{x}') = \exp\left(-\gamma|\mathbf{x}-\mathbf{x}'|^2\right),
\end{equation}
where $\mathbf{x}$ and $\mathbf{x}'$ represent two feature vectors, and $\gamma$ denotes the kernel scale parameter, which was tuned independently for each witness model using the validation dataset. However, the model’s performance did not exhibit a significant change while changing $\gamma$. The consistent high performance across all three classification tasks confirms the effectiveness of RBF kernel for entanglement classification problems. Our final models achieved the test accuracies 96\%,  98\%, and 96\%, for the models $\mathcal{M}_{GHZ}$, $\mathcal{M}_{W}$, and $\mathcal{M}_{S}$ respectively. The confiusion matrices of all three models are presented in Figure \ref{fig:confiusionMatrix}. When considering the sequential application of all three models for the classification task of an arbitrary state (see Figure \ref{fig:modelStructure}), the overall system accuracy reaches 95\%. The confusion matrix of the cascaded model is represented in Figure \ref{fig:CMHoleModel}. 

To enhance model reliability, particularly for practical implementations, we employ two other key strategies: (1) systematically evaluating out-of-distribution (OOD) samples to assess generalization beyond the current data distribution, and (2) augmenting the training data with controlled noise to improve robustness. This approach shows the strength of our models in real-world experimental conditions.

\subsection{Performance for Out-Of-Distribution dataset}
For evaluating model performance on Out-Of-Distribution (OOD) data, we consider four families of three-qubit states with analytically characterized entanglement properties. This set  includes three distinct classes of positive partial transpose entangled states (PPTes) \cite{acinClassification3q, horodecki, UPB}, along with the X-states \cite{XState}. The PPT-entangled states comprise the following families:\\

(i) Horodecki's $2 \otimes 4$ bound entangled states~\cite{horodecki}, represented by the following density matrix
\begin{equation}
	\label{eq:horodecki}
	\rho_{H} = \frac{1}{7a+1}
	\begin{bmatrix}
		a & 0 & 0 & 0 & 0 & a & 0 & 0 \\
		0 & a & 0 & 0 & 0 & 0 & a & 0 \\
		0 & 0 & a & 0 & 0 & 0 & 0 & a \\
		0 & 0 & 0 & a & 0 & 0 & 0 & 0 \\
		0 & 0 & 0 & 0 & \dfrac{1+a}{2} & 0 & 0 & \dfrac{\sqrt{1-a^2}}{2} \\
		a & 0 & 0 & 0 & 0 & a & 0 & 0 \\
		0 & a & 0 & 0 & 0 & 0 & a & 0 \\
		0 & 0 & a & 0 & \dfrac{\sqrt{1-a^2}}{2} & 0 & 0 & \dfrac{1+a}{2}
	\end{bmatrix}.
\end{equation}
Where, $0 < a < 1$.\\

(ii) Edge states~\cite{acinClassification3q}, which are represented by the density matrix
\begin{equation}
\rho_{E} = \frac{1}{n}
\begin{bmatrix}
1 & 0 & 0 & 0 & 0 & 0 & 0 & 1 \\
0 & a & 0 & 0 & 0 & 0 & 0 & 0 \\
0 & 0 & b & 0 & 0 & 0 & 0 & 0 \\
0 & 0 & 0 & c & 0 & 0 & 0 & 0 \\
0 & 0 & 0 & 0 & c^{-1} & 0 & 0 & 0 \\
0 & 0 & 0 & 0 & 0 & b^{-1} & 0 & 0 \\
0 & 0 & 0 & 0 & 0 & 0 & a^{-1} & 0 \\
1 & 0 & 0 & 0 & 0 & 0 & 0 & 1
\end{bmatrix}
\label{eq:edge},
\end{equation}
with $a, b, c > 0$ and $n = 2 + a + 1/a + b + 1/b + c + 1/c$.\\

(iii) The states generated via unextendible product bases (UPB)~\cite{UPB}. They are locally rotated versions of the state 
\begin{equation}
\label{eq:UPB}
\rho_{\text{UPB}} = 1 - \sum_{i=1}^4 |v_i\rangle \langle v_i|,
\end{equation}
where $|v_1\rangle = |000\rangle$, $|v_2\rangle = |-+1\rangle$, $|v_3\rangle = |+1-\rangle$ and $|v_4\rangle = |1-+\rangle$.\\

All the above three sets are entangled states that belong to the bi-separable set, i.e. all belong to $B \setminus S$. Hence, it is expected that $\mathcal{M}_{GHZ}$ and $\mathcal{M}_{W}$ label them as 0, and $\mathcal{M}_{B}$ labels them as 1. \\

The fourth family is the set of X-states, which are defined by having non-zero values exclusively on their diagonal and anti-diagonal elements.
For X-states, based on the analytical solution in \cite{Seyed2012}, the dataset divided into two parts:  Genuine Multipartite Entanglement (GME) and non-GME states. Since $W$-type entanglement are rare in comparison to $GHZ$-type entanglement \cite{Wolfgang2000}, almost all randomly generated GME X-states fall into the $GHZ$ class. Therefore, it is reasonable to conclude that only $\mathcal{W}_{GHZ}$ is meaningful for evaluating this set. It is expected that $\mathcal{W}_{GHZ}$ labaled GME state as 1 and non-GME as 0. Accordingly, in Table \ref{tab:ood_result}, the entries corresponding to $\mathcal{W}_{W}$ and $\mathcal{W}_{B}$ for X-states are indicated by “--”, as no meaningful results can be obtained for them.

\begin{table}[ht]
\centering
\caption{Classification accuracy (\%) of witness models on out-of-distribution sets}
\begin{tabular}{lcccc}
\toprule
Model & Horodecki States &  Edge State & UPB States & X-State \\
\midrule
$\mathcal{W}_{B}$ & 99.9\% & 76\% & 99.9\% &  -- \\
$\mathcal{W}_{W}$ & 93\% & 98\% & 99.9\% &  -- \\
$\mathcal{W}_{GHZ}$ & 99.9\% & 99.9\% & 99.9\% & 98\% \\
\bottomrule
\label{tab:ood_result}
\end{tabular}
\end{table}

\subsection{Model performance under noisy conditions}
In order to check model robustness under realistic conditions that involve noise effects, we have evaluated model performance by using noisy dataset. This dataset has the same structure as the one which was described in section \ref{sec:datasetGeneration}, in addition with noise effects. Our noise model incorporates three fundamental quantum noise channels as follow:

\begin{itemize}
    \item \textbf{Amplitude Damping}, which simulates energy dissipation and represents qubit relaxation.
    
    \item \textbf{Phase Damping}, which simulates loss of quantum phase information without energy loss.
 
    \item \textbf{Depolarizing}, which represents destroying quantum information of states.
\end{itemize}

\begin{figure}[h]
    \centering
    \includegraphics[scale=0.5]{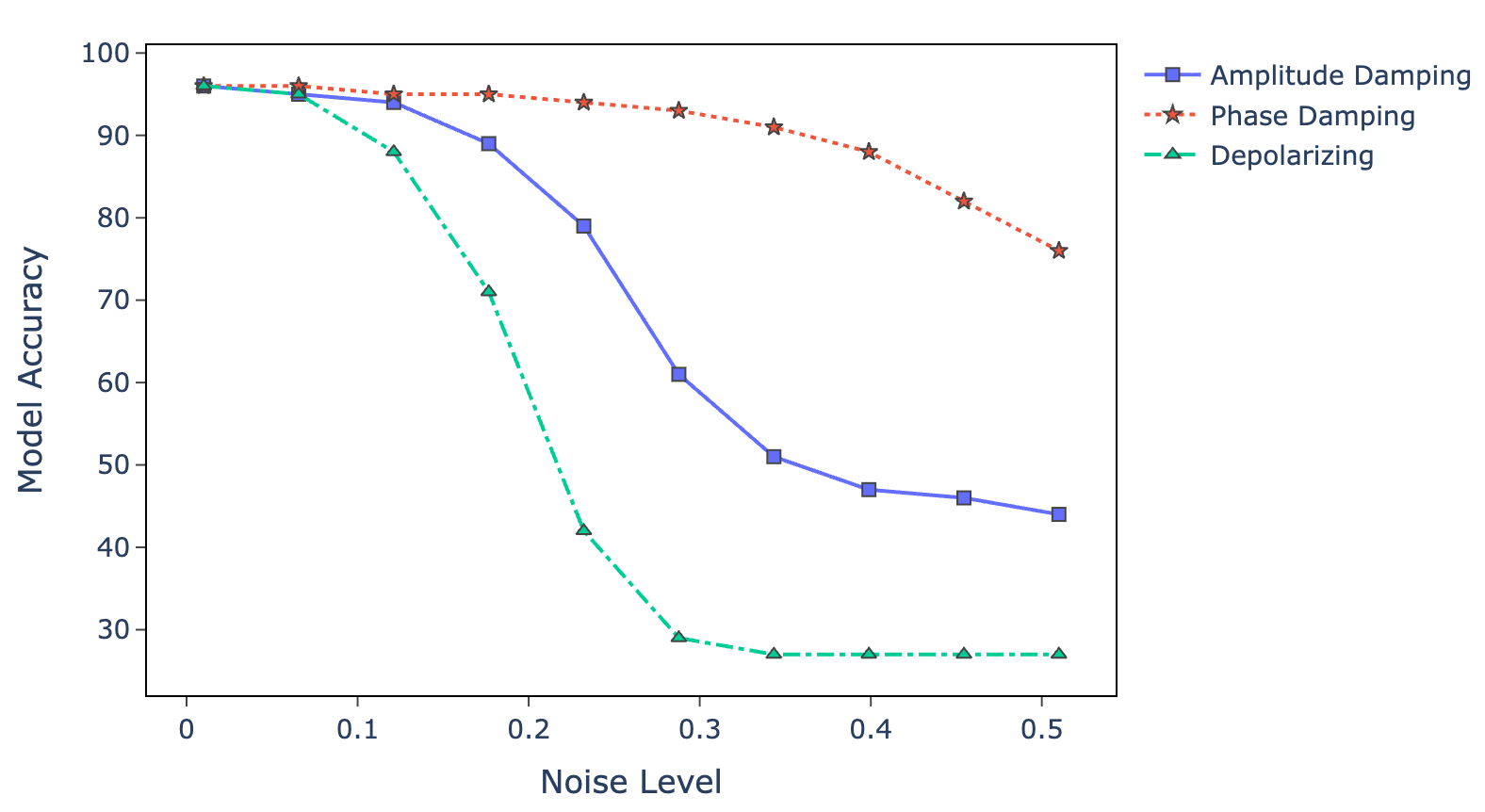}
    \caption{Model accuracy versus noise strength, for different noise channels applied independently to all three qubits.}
    \label{fig:noiseEffect}
\end{figure}

This analysis was implemented using the noise module of the PennyLane quantum machine learning library~\cite{bergholm2018pennylane}, which provides built-in noise channels that accurately model relevant physical processes. Noise with strengths ranging from 0.01 to 0.5 was applied independently to all three qubits of each quantum state in the dataset, and the resulting classification accuracy over the four entanglement classes was evaluated at each noise level. Figure~\ref{fig:noiseEffect} summarizes the model accuracy as a function of noise strength for the different noise channels considered. The results show that the model is most robust against phase-damping and most susceptible to depolarizing. Notably, for all noise types considered, the classification performance remains largely unaffected for noise strengths below 0.1.

\section{Deriving the optimal model}
\label{sec:optimalWitness}
 
Beyond constructing entanglement classification models, an essential step toward experimental feasibility is model optimization. As shown in Sec.~\ref{sec:datasetGeneration}, a complete  tomography of a three-qubit state requires 63 independent parameters, corresponding to 63 independent measurements. Full state tomography is therefore experimentally demanding, since each projective measurement collapses the quantum state and requires repeated state preparation. This renders the acquisition of all observables resource-intensive and often impractical. To address this challenge, we introduce a model-optimization approach that significantly reduces the measurement overhead while retaining the classification performance.

The core of this approach is the construction of witness models with varying numbers of measured observables, ranging from 3 to the complete set of 63. Each model is defined by a selected subset of observables, which serve as features for classifying an arbitrary quantum state into one of the four entanglement classes. As a representative example, one may consider two models based on feature sets of cardinalities $m$ and $n$, with $2<m<n<64$. In this case, the model employing the larger feature set exhibits greater expressive capacity and, consequently, improved classification performance.

To reduce the number of required measurements—or, in model terms, the number of features— a systematic feature selection protocol based on feature importance analysis is implemented. Our approach evaluates the contribution of each feature to model performance through an iterative procedure: for each feature, first change the original value of the feature with a random value, second evaluate the model's performance degradation, and third, repeat this process 50 times to obtain statistically reliable importance estimates. The importance metric is directly quantified by Equation \ref{eq:rate}, where $R_{f_i}$, denotes $i^{th}$ feature importance rate, $N = 50$ is the total number of iterations, $S$ represents baseline model performance and $S_j^{(i)}$ indicates model performance after randomazing $i^{th}$ feature in the $j^{th}$ iteration.

\begin{equation}
R_{f_i} = \frac{1}{N} \sum_{j=1}^{N} S - S_j^{(i)}.
\label{eq:rate}
\end{equation}

\begin{figure}[h]
    \centering
    \includegraphics[scale=0.5]{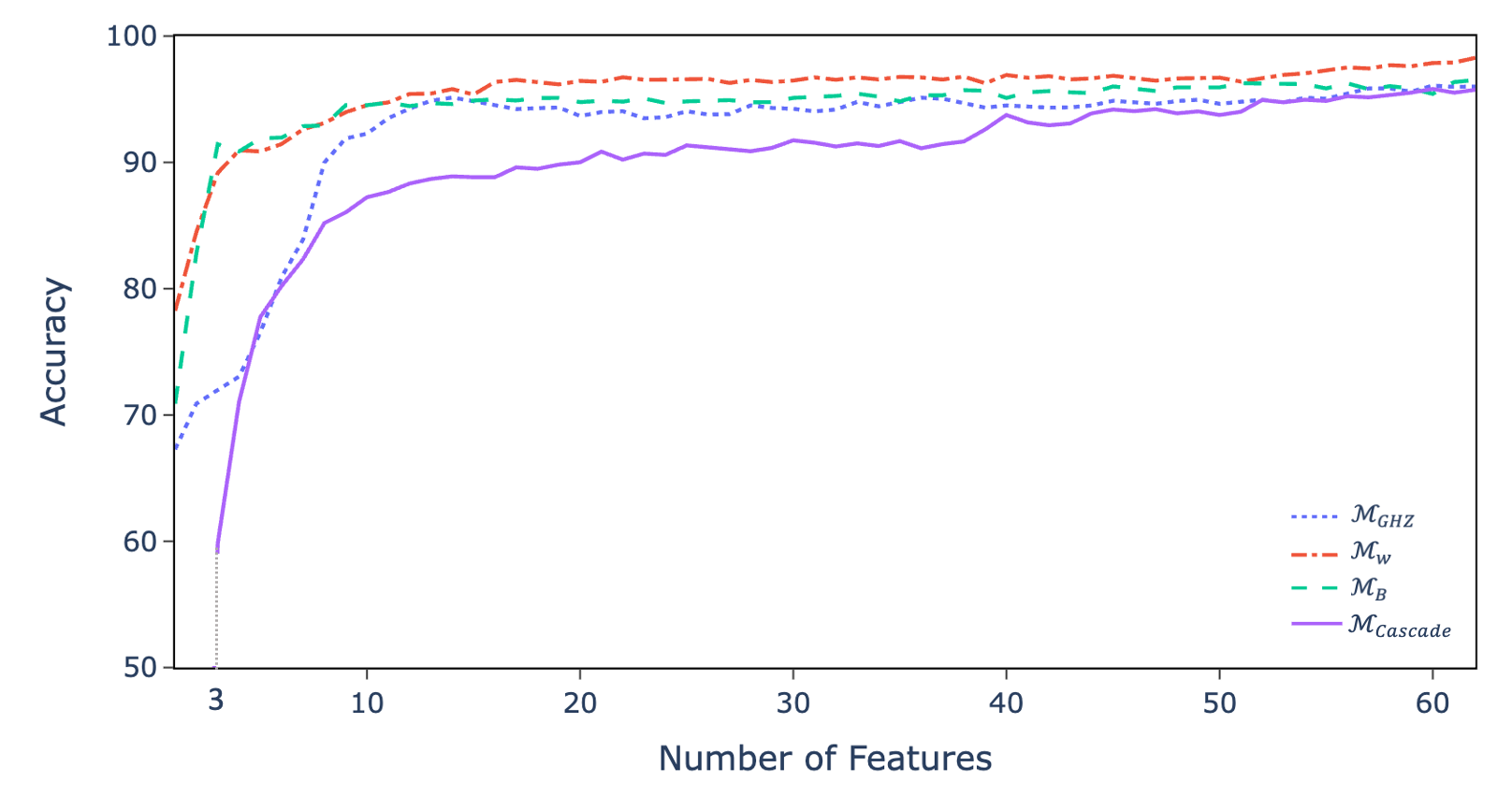}
    \caption{Classification accuracy of models $\mathcal{M}_B$, $\mathcal{M}_W$,  $\mathcal{M}_{GHZ}$, and $\mathcal{M}_{\mathrm{cascade}}$, as a function of the number of retained features. Each curve is obtained by sequentially adding features according to their ranked importance.}
    \label{fig:accuracy_per_features}
\end{figure}

Using this method, we first assign importance scores to all features of the three models, $\mathcal{M}_B$, $\mathcal{M}_W$, and $\mathcal{M}_{GHZ}$, and rank them accordingly. For each model, we then construct a sequence of 63 distinct models that differ only in the number of retained features: the model with a single feature includes only the highest-ranked feature, while the model with $k$ features includes the top $k$ features in the ordered list. Figure~\ref{fig:accuracy_per_features} displays the classification accuracy as a function of the number of retained features, showing that the accuracy saturates when approximately 20 features are included, with additional features yielding negligible improvement. Notably, high classification performance---$93\%$ for $\mathcal{M}_{GHZ}$, $95\%$ for $\mathcal{M}_W$, and $94\%$ for $\mathcal{M}_B$---is achieved by using fewer than $30\%$ of the total features, corresponding to a substantial reduction in the number of required measurements and, consequently, in the experimental overhead.

Until now, feature reduction has been performed separately for the three models $\mathcal{M}_B$, $\mathcal{M}_W$, and $\mathcal{M}_{GHZ}$. Since the final model integrates these three ones, it is crucial to assess feature rankings and their contributions within the combined model. A natural approach is to derive a consensus ranking by merging the individually ranked feature lists. To implement this efficiently, we introduce the following algorithm for feature aggregation across the models.

\begin{figure}[h]
    \centering
    \includegraphics[scale=0.5]{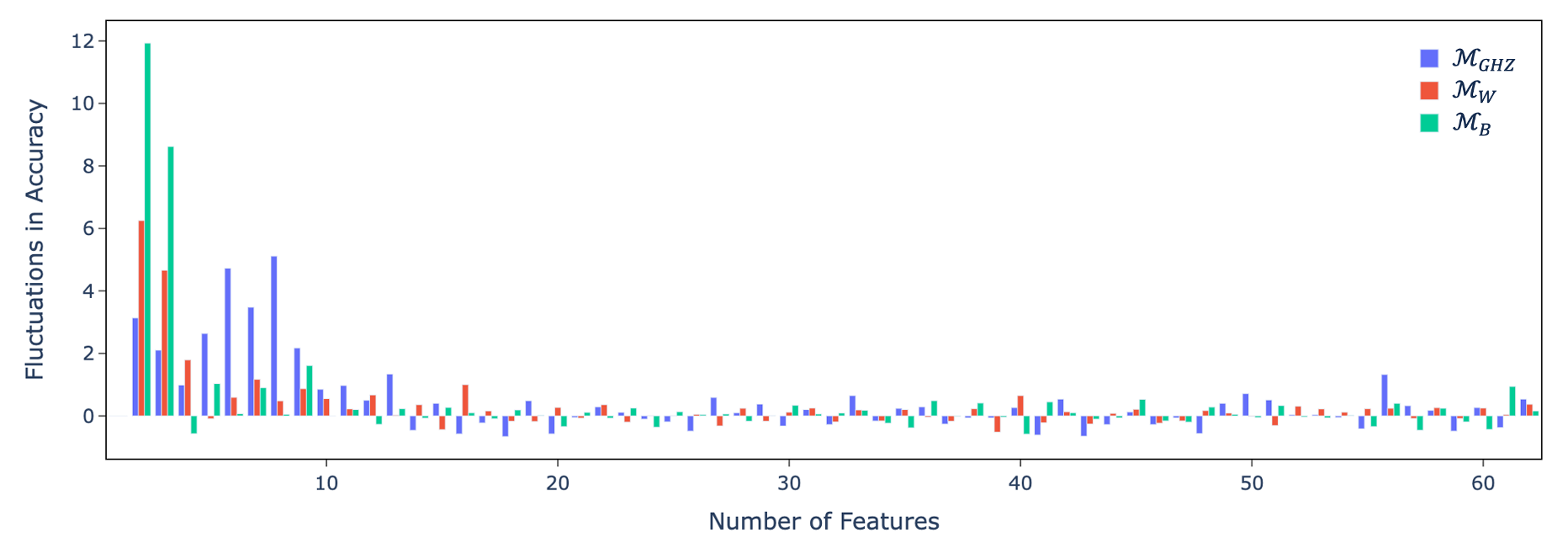}
    \caption[Accuracy variations of each model with the addition of new features]{Accuracy variations of each model as new features are added. The figure contains 62 groups of three columns, corresponding (left to right) to $\mathcal{M}_{GHZ}$, $\mathcal{M}_W$, and $\mathcal{M}_B$. The $i$‑th group shows accuracy after adding the $(i+1)$‑th most important feature. According to this plot, constructing a cascaded model with four features requires including the second-ranked feature of $\mathcal{M}_B$ alongside the top features of the three individual models.

}
    \label{fig:fluctuations_in_accuracy}
\end{figure}

\paragraph{Feature Consensus Algorithm:}
This algorithm provides a ranked list of features for the cascaded model by comparing features of the same rank across the three models. The following steps describe how this algorithm works:

\begin{itemize}[leftmargin=*,noitemsep,topsep=0pt]

    \item Initializing- The most important features of the models $\mathcal{M}_B$, $\mathcal{M}_{W}$, and $\mathcal{M}_{GHZ}$ were selected as first three features of the cascade model. Hence, the cascade model starts by exploiting at least 3 features. 
    
    \item Selecting $i$-th feature- To select the next features for the cascade models, we evaluate the accuracy increase for the models $\mathcal{M}_B$, $\mathcal{M}_W$, and $\mathcal{M}_{GHZ}$ after adding the $i$-th feature. Figure \ref{fig:fluctuations_in_accuracy} illustrates this accuracy variations for all 63 features. We then compare and sort the results of the $i$-th features (i.e. $i$-th set of three columns in Figure \ref{fig:fluctuations_in_accuracy}). The top feature among  them is selected, and the other two ones are added to a separate list called the reserve list. If the top feature was previously selected, then the first non-repeated feature is selected from the beginning of the reserve list.
    
    \item Ending- Continue the previous step until all features are selected.
\end{itemize}

The rationale behind this selection is consistent with the interpretation that if a feature contributes the highest accuracy increase in its own model, incorporating it into the integrated  model—which combines all three models—will lead to a greater overall accuracy improvement. The purple-solid line in Figure \ref{fig:accuracy_per_features} represents the accuracy variations of the final model based on the sorted feature list derived from the above proposed algorithm. Table \ref{tab:feature_ablation} represents the list of sorted features.

Based on the model accuracies shown in Figure~\ref{fig:accuracy_per_features} for $\mathcal{M}_{\mathrm{cascade}}$, only the first 30 features listed in Table~\ref{tab:feature_ablation} are sufficient to classify an arbitrary quantum state into one of the four labels with $91\%$ accuracy. This demonstrates that the classification task can be performed reliably using less than half of the original measurement set. Accordingly, this feature selection strategy enables a substantial reduction in experimental resource requirements.

\section{Conclusion}
\label{sec:discussion}

This study demonstrates the effectiveness of classical machine learning, specifically Support Vector Machines (SVMs), in classifying entanglement across the full state space of three-qubit systems. Unlike deep learning approaches, which often rely on complex and less interpretable architectures, SVMs achieve high performance while accurately capturing the underlying structure of the problem. The proposed framework attains near-perfect classification, reaching up to $95\%$ accuracy, and allows a tunable measurement strategy with 3 to 63 features, yielding accuracies between $71\%$ and $95\%$. For instance, using only 10 features (approximately $16\%$ of full state tomography), the model correctly classifies over $87\%$ of states, while 20 features suffice to surpass $90\%$ accuracy. These results underscore the framework’s efficiency in feature reduction and practical implementation. Compared to existing methods, this approach offers three main advantages: (i) higher classification accuracy, (ii) substantial reduction in required measurements, and (iii) coverage of a broader range of mixed three-qubit states.
For instance, our method achieves high classification accuracy across the full space of three-qubit mixed states, extending beyond the scope of previous SVM-based approaches. In \cite{sanavio2023}, SVMs reached 95\% accuracy on mixed states, while \cite{greenwoodML} was limited to Werner-like GHZ and W states. Compared to deep learning-based methods, our approach also demonstrates competitive performance: multi-layer perceptron models in \cite{julioDNN} achieved 90\% accuracy, and the method in \cite{naema2023} reached 80\% for three-qubit states. Convolutional neural networks in \cite{jaroSemiSupervised} reported 96\% average accuracy for detecting entangled states with positive partial transpose, whereas our framework achieves up to 99.9\% in similar benchmarks. These comparisons highlight the broader applicability and high performance of our SVM-based classification framework.

We finally note that, SVMs provide a systematic framework for constructing entanglement witness operators. In some cases, the optimized separating hyperplane can be mapped to a corresponding witness, akin to the approach in \cite{Havlicek2019}. However, this mapping is not always feasible due to the irreversibility of certain kernel functions, and its detailed application to the present study is reserved for a future work.

\begin{table}[ht]
\caption{Gradual improvement of the accuracies (\%) of each individual models $\mathcal{M}_B$, $\mathcal{M}_W$, $\mathcal{M}_{GHZ}$, and the final cascaded model $\mathcal{M}_{cascade}$. Sequential addition of features is based on their importance-ranking derived for the final model.} 
\centering
\begin{minipage}{0.48\textwidth}
\centering
\small
\setlength{\tabcolsep}{3pt}
\begin{tabular}{c c c c c}
\hline
\textbf{Features} & $\mathcal{M}_{B}$ & $\mathcal{M}_{W}$ & $\mathcal{M}_{GHZ}$ & \textbf{Cascade Model} \\
\hline
\ IIZ & $70.8$ & $78.2$ & $67.2$ & -- \\
+ ZZY & $82.8$ & $84.5$ & $70.9$ & -- \\
+ YZY & $91.4$ & $89.1$ & $71.9$ & $52.2$ \\
+ IIY & $90.8$ & $90.9$ & $73.0$ & $52.9$ \\
+ IIX & $91.9$ & $90.8$ & $80.9$ & $59.8$ \\
+ XZX & $91.9$ & $91.4$ & $83.8$ & $71.0$ \\
+ XYY & $92.8$ & $92.6$ & $89.9$ & $77.7$ \\
+ YXX & $92.9$ & $93.1$ & $91.8$ & $80.2$ \\
+ XZI & $94.5$ & $93.9$ & $92.2$ & $82.3$ \\
+ YZI & $94.5$ & $94.5$ & $93.4$ & $85.1$ \\
+ IYI & $94.7$ & $94.7$ & $94.2$ & $86.0$ \\
+ IXI & $94.4$ & $95.4$ & $94.8$ & $87.2$ \\
+ XYZ & $94.6$ & $95.4$ & $95.1$ & $87.6$ \\
+ ZXY & $94.6$ & $95.8$ & $94.8$ & $88.3$ \\
+ ZIY & $94.8$ & $95.3$ & $94.5$ & $88.6$ \\
+ ZXX & $94.9$ & $96.3$ & $94.2$ & $88.8$ \\
+ YZX & $94.9$ & $96.5$ & $94.2$ & $88.8$ \\
+ IXX & $95.0$ & $96.3$ & $94.3$ & $88.8$ \\
+ XYI & $95.1$ & $96.1$ & $93.6$ & $89.6$ \\
+ YIZ & $94.7$ & $96.4$ & $93.9$ & $89.4$ \\
+ XIX & $94.8$ & $96.3$ & $94.0$ & $89.8$ \\
+ YYY & $94.8$ & $96.7$ & $93.4$ & $90.0$ \\
+ XIZ & $95.0$ & $96.5$ & $93.5$ & $90.8$ \\
+ IZY & $94.6$ & $96.5$ & $94.0$ & $90.2$ \\
+ IYX & $94.8$ & $96.5$ & $93.7$ & $90.6$ \\
+ YXY & $94.8$ & $96.6$ & $93.8$ & $90.5$ \\
+ YZZ & $94.9$ & $96.2$ & $94.5$ & $91.3$ \\
+ XXZ & $94.7$ & $96.5$ & $94.3$ & $91.1$ \\
+ XXX & $94.7$ & $96.3$ & $94.2$ & $90.9$ \\
+ IYY & $95.1$ & $96.4$ & $94.0$ & $90.8$ \\
+ XIY & $95.1$ & $96.7$ & $94.1$ & $91.1$ \\
+ ZYZ & $95.2$ & $96.5$ & $94.7$ & $91.7$ \\

\hline
\end{tabular}
\end{minipage}
\hfill
\begin{minipage}{0.48\textwidth}
\centering
\small
\setlength{\tabcolsep}{3pt}
\begin{tabular}{c c c c c}
\hline
\textbf{Features} & $\mathcal{M}_{B}$ & $\mathcal{M}_{W}$ & $\mathcal{M}_{GHZ}$ & \textbf{Cascade Model} \\
\hline
+ YYZ & $95.4$ & $96.7$ & $94.4$ & $91.5$ \\
+ YIX & $95.2$ & $96.5$ & $94.7$ & $91.2$ \\
+ ZYY & $94.8$ & $96.7$ & $95.1$ & $91.5$ \\
+ ZZX & $95.3$ & $96.5$ & $95.0$ & $91.2$ \\
+ IXZ & $95.3$ & $96.7$ & $94.7$ & $91.6$ \\
+ ZZZ & $95.7$ & $96.7$ & $94.3$ & $91.1$ \\
+ IXY & $95.6$ & $96.7$ & $94.5$ & $91.4$ \\
+ YYX & $95.1$ & $96.5$ & $94.3$ & $91.6$ \\
+ XXI & $95.5$ & $96.7$ & $94.3$ & $92.6$ \\
+ IZZ & $95.6$ & $96.2$ & $94.3$ & $93.7$ \\
+ ZIZ & $95.5$ & $96.9$ & $94.5$ & $93.1$ \\
+ XZY & $95.4$ & $96.7$ & $94.8$ & $92.9$ \\
+ YYI & $96.0$ & $96.8$ & $94.8$ & $93.0$ \\
+ ZXI & $95.8$ & $96.5$ & $94.6$ & $93.8$ \\
+ IZI & $95.6$ & $96.6$ & $94.8$ & $94.1$ \\
+ XYX & $95.9$ & $96.8$ & $94.9$ & $94.0$ \\
+ ZYI & $95.9$ & $96.6$ & $94.6$ & $94.2$ \\
+ XII & $95.9$ & $96.4$ & $94.8$ & $93.8$ \\
+ IZX & $96.2$ & $96.6$ & $94.9$ & $94.0$ \\
+ ZXZ & $96.2$ & $96.7$ & $94.7$ & $93.7$ \\
+ YIY & $96.1$ & $96.7$ & $95.1$ & $94.0$ \\
+ ZZI & $96.2$ & $96.9$ & $95.0$ & $94.9$ \\
+ XXY & $95.8$ & $97.0$ & $95.3$ & $94.7$ \\
+ ZIX & $96.2$ & $97.2$ & $95.8$ & $94.6$ \\
+ YXZ & $95.8$ & $97.5$ & $95.8$ & $94.8$ \\
+ ZII & $96.0$ & $97.4$ & $95.6$ & $95.2$ \\
+ YII & $95.8$ & $97.6$ & $96.0$ & $95.1$ \\
+ YXI & $95.4$ & $97.8$ & $95.9$ & $95.4$ \\
+ XZZ & $96.3$ & $97.9$ & $96.0$ & $95.5$ \\
+ IYZ & $96.5$ & $98.2$ & $96.0$ & $95.3$ \\
+ ZYX & $96.2$ & $98.3$ & $96.1$ & $95.6$ \\
\  & \  & \  & \  & \  \\
\hline
\end{tabular}
\end{minipage}
\label{tab:feature_ablation}
\end{table}

\clearpage
\bibliographystyle{ieeetr}
\bibliography{references}

@article{acinClassification3q,
  title={Classification of mixed three–qubit states},
  author={A. Acin and D. Brub and M. Lewenstein and A. Sanpera},
  journal={Physical Review Letters},
  volume={87},
  year={2001}
}

@article{brubCharEnt,
  title={Characterizing Entanglement},
  author={D. Brub},
  journal={Journal of Mathematical Physics},
  year={2001}
}

@article{chenUnsupervise,
  title={Detecting quantum entanglement with unsupervised learning},
  author={Y. Chen and Y. Pan and G. Zhang and S. Cheng},
  journal={Quantum Science and Technology},
  volume={7},
  year={2022}
}

@article{PPT,
	title={Separability of Mixed States: Necessary and Sufficient Conditions},
	author={M. Horodecki and P. Horodecki and R. Horodecki and K. Horodecki},
	journal={Physics Letters A},
	volume={223},
	year={1996}
}

@article{julioDNN,
  title={Entanglement detection with classical deep neural networks},
  author={J. Urena and A. Sojo and J. Bermejo-Vega and D. Manzano},
  journal={Scientific Reports},
  volume={14},
  year={2024}
}

@article{jaroSemiSupervised,
  title={Identification of quantum entanglement with Siamese convolutional NNs and semi-supervised learning},
  author={J. Pawłowski and M. Krawczyk},
  journal={Phys. Rev. Applied},
  volume={22},
  year={2024}
}

@article{greenwoodML,
  title={Machine Learning Derived Entanglement Witnesses},
  author={A. C. B. Greenwood and L. T. H. Wu and E. Y. Zhu and B. T. Kirby and L. Qian},
  journal={Phys. Rev. Applied},
  volume={19},
  year={2023}
}

@article{MADistribution,
  title={Improving application performance with biased distributions of quantum states},
  author={S. Lohani and J. M. Lukens and D. E. Jones and T. A. Searles and R. T.
Glasser and B. T. Kirby},
  journal={Phys. Rev. Research},
  volume={3},
  year={2021}
}

@article{UPB,
  title={Unextendible product bases and bound entanglement},
  author={C. H. Bennett and D.P. DiVincenzo and T. Mor and P.W. Shor and J.A. Smolin and B.M. Terhal},
  journal={Phys Rev Lett},
  volume={82},
  year={1999}
}

@article{horodecki,
  title={Separability criterion and inseparable mixed states with positive partial transposition},
  author={P. Horodecki},
  journal={Phys Lett A},
  volume={232},
  year={1997}
}

@article{XState,
  title={Generalized X states of N qubits and their symmetries},
  author={S. Vinjanampathy and A. R. P. Rau},
  journal={Phys Lett A},
  volume={232},
  year={2010}
}

@article{Seyed2012,
  author       = {S. M. H. Rafsanjani and M. Huber and C. J. Broadbent and J. H. Eberly},
  title        = {Genuinely Multipartite Concurrence of N-Qubit X-Matrices},
  journal      = {Phys Rev A},
  volume       = {86},
  number       = {6},
  pages        = {062307},
  year         = {2022},
  doi          = {10.1103/PhysRevA.86.062307}
}

@article{Wolfgang2000,
  author       = {W. Dür and D. Kafri and I. Cirac},
  title        = {Three Qubits Can Be Entangled in Two Inequivalent Ways},
  journal      = {Phys. Rev. A},
  volume       = {62},
  number       = {6},
  pages        = {062314},
  year         = {2000},
  doi          = {10.1103/PhysRevA.62.062314}
}

@article{bergholm2018pennylane,
  author       = {V. Bergholm and J. Izaac and M. Schuld and C. Gogolin and S. Ahmed and V. Ajith and M. S. Alam and G. Alonso-Linaje and B. A. Narayanan and A. Asadi and {others}},
  title        = {PennyLane: Automatic Differentiation of Hybrid Quantum-Classical Computations},
  journal      = {arXiv preprint arXiv:1811.04968},
  year         = {2018},
  url          = {https://arxiv.org/abs/1811.04968}
}

@article{sanavio2023,
  author       = {C. Sanavio and E. Tignone and E. Ercolessi},
  title        = {Entanglement Classification via Witness Operators Generated by Support Vector Machine},
  journal      = {Eur. Phys. J. Plus},
  volume       = {138},
  number       = {2},
  pages        = {109},
  year         = {2023},
  doi          = {10.1140/epjp/s13360-023-03689-9}
}

@article{naema2023,
  author       = {N. Asif and U. Khalid and A. Khan and T. Q. Duong and H. Shin},
  title        = {Entanglement Detection With Artificial Neural Networks},
  journal      = {Sci. Rep.},
  volume       = {13},
  number       = {1},
  pages        = {309},
  year         = {2023},
  doi          = {10.1038/s41598-022-26606-z}
}

@article{gharibian2010strong,
  author       = {S. Gharibian},
  title        = {Strong NP-Hardness of the Quantum Separability Problem},
  journal      = {Quantum Inf. Comput.},
  volume       = {10},
  number       = {3-4},
  pages        = {343--360},
  year         = {2010},
  url          = {https://arxiv.org/abs/0810.4507}
}

@article{bennett1996concentrating,
  author       = {C. H. Bennett and H. J. Bernstein and S. Popescu and B. Schumacher},
  title        = {Concentrating Partial Entanglement by Local Operations},
  journal      = {Phys. Rev. A},
  volume       = {53},
  number       = {4},
  pages        = {2046--2052},
  year         = {1996},
  doi          = {10.1103/PhysRevA.53.2046}
}

@article{chitambar2019quantum,
  author       = {E. Chitambar and G. Gour},
  title        = {Quantum Resource Theories},
  journal      = {Rev. Mod. Phys.},
  volume       = {91},
  number       = {2},
  pages        = {025001},
  year         = {2019},
  doi          = {10.1103/RevModPhys.91.025001}
}

@article{roik2023physrev,
  author       = {V. Trávníček and J. Roik and K. Bartkiewicz and A. Černoch and P. Horodecki and K. Lemr},
  title        = {Sensitivity versus selectivity in entanglement detection via collective witnesses},
  journal      = {Phys. Rev. Research},
  volume       = {6},
  year         = {2024},
  doi          = {10.1103/PhysRevResearch.6.033056}
}

@book{svm,
  author       = {T. Hastie and R. Tibshirani and J. Friedman},
  title        = {The Elements of Statistical Learning: Data Mining, Inference, and Prediction},
  edition      = {2nd},
  year         = {2009},
  publisher    = {Springer},
  address      = {New York, NY, USA}
}

@article{polson2011data,
  title={Data augmentation for support vector machines},
  author={N. Polson and J. G. Scott},
  journal={Bayesian Analysis},
  volume={6},
  number={1},
  pages={1--24},
  year={2011},
  publisher={Project Euclid}
}

@article{siewert2011,
  title={Separability criteria for mixed three-qubit states},
  author={J. Siewert and A. Elben},
  journal={Physical Review A},
  volume={83},
  number={6},
  pages={062337},
  year={2011},
  publisher={APS},
  doi={10.1103/PhysRevA.83.062337}
}

@article{Eltschka2008,
	author={C. Eltschka and A. Osterloh and J. Siewert and A. Uhlmann},
	title= {Three-tangle for mixtures of generalized GHZ and generalized W states},
	journal={New J. Phys},
	volume={10},
	year={2008}
}

@misc{Mahdian2025,
	author= {M. Mahdian and Z. Mousavi},
	title= {Optimal Entanglement Witness of Multipartite Systems Using Support Vector Machine Approach},
	year= {2025},
	eprint= {2504.18163},
	archivePrefix= {arXiv},
	primaryClass = {quant-ph},
	url= {https://arxiv.org/abs/2504.18163}
}

@article{Havlicek2019,
	author= {V. Havlicek and A. D. Corcoles and K Temme and A. W. Harrow and A. Kandala and J. M. Chow and J. M. Gambetta },
	title= {Supervised learning with quantum-enhanced feature spaces},
	journal={Nature},
	volume={567},
	year={2019}
}

@article{Sabın2008,
  title={A classification of entanglement in three-qubit systems},
  author={C. Sabin and G. Garca-Alcaine},
  journal={The European Physical Journal D},
  volume={48},
  year={2008},
  doi={10.1140/epjd/e2008-00112-5}
}

@article{Borsten2009,
  title={Freudenthal triple classification of three-qubit entanglement},
  author={L. Borsten and D. Dahanayake and M. J. Duff and W. Rubens and H. Ebrahim},
  journal={Phys. Rev. A},
  volume={80},
  year={2009},
  doi={0.1103/PhysRevA.80.032326}
}

@article{Prakash2011,
  title={Quantum teleportation using entangled 3-qubit states and the magic bases},
  author={H. Prakash and A. K. Maurya},
  journal={Optics Communications},
  volume={284},
  year={2011},
  doi={10.1016/j.optcom.2011.07.002}
}

@article{Gühne2009,
  title={Entanglement detection},
  author={O. Gühne and G. Toth},
  journal={Physics Reports},
  volume={474},
  year={2009},
  doi={10.1016/j.physrep.2009.02.004}
}

@article{Michal1996,
  title={Separability of Mixed States: Necessary and Sufficient Conditions},
  author={M. Horodecki and P. Horodecki, R. Horodecki},
  journal={Physics Letters A},
  volume={223},
  year={1996},
  doi={10.1016/S0375-9601%2896%2900706-2}
}

@inproceedings{Gurvits2003,
  author    = {L. Gurvits},
  title     = {Classical Deterministic Complexity of Edmonds' Problem and Quantum Entanglement},
  booktitle = {Proceedings of the Thirty-Fifth Annual ACM Symposium on Theory of Computing},
  pages     = {10--19},
  year      = {2003},
  publisher = {ACM},
  doi       = {10.1145/780542.780545}
}

@article{Liu2022,
  title={Fundamental Limitation on the Detectability of Entanglement},
  author={P. Liu and Z. Liu and S. Chen and X. Ma},
  journal={Phys. Rev. Lett},
  volume={129},
  year={2022},
  doi={10.1103/PhysRevLett.129.230503}
}

\end{document}